\documentstyle[aps,twocolumn,epsfig]{revtex}
\newtheorem{theorem}{Theorem}
\newtheorem{acknowledgement}[theorem]{Acknowledgement}

\begin{document}

\twocolumn[\hsize\textwidth\columnwidth\hsize\csname 
@twocolumnfalse\endcsname

\title{Stripes disorder and correlation lengths in doped antiferromagnets}
\author{Oron Zachar}
\address{ICTP, 11 strada Costiera, Trieste 34100, Italia}
\date{\today}
\maketitle

\widetext
\begin{abstract}

For stripes in doped antiferromagnets, we find that the ratio of spin and charge 
correlation lenghts, $\xi _{s}/\xi _{c}$, provide a sharp criterion for determining 
the dominant form of disorder in the system. If stripes are disordered predominantly 
by topological defects then $\xi _{s}/\xi _{c}\lesssim 1$. 
In contast, if stripes correlations are disordered primarily by non-topological 
elastic deformations (i.e., a Bragg-Glass type of disorder) then 
$1<\xi _{s}/\xi _{c}\lesssim 4$ is expected. Therefore, the observation of 
$\xi _{s}/\xi _{c}\approx 4$ in $(LaNd)_{2-x}Sr_{x}CuO_{4}$ and 
$\xi _{s}/\xi _{c}\approx 3$ in $La_{2/3}Sr_{1/3}NiO_{4}$ invariably implies 
that the stripes are in a Bragg glass type state, and topological defects are 
much less relevant than commonly assumed. Expected spectral properties 
are discussed. Thus, we establish the basis for any theoretical analysis of the 
experimentally obsereved glassy state in these material.

\smallskip
\end{abstract}

 ]

\narrowtext

\section{Introduction}

Recent experiments\cite%
{Tranquada99-(LaNd)SrCuO-glass,Hammel2000-LaEuSrCuO-glass,Hammel99-NiO-glass,Cheong2000-NiO.133,Yamada2000-LSCO-underdoped}
shed new light and conjure up new questions on the glassy state of doped
antiferromagnets. There is growing experimental evidence and theoretical
arguments that electronic charge and spin ''topological stripes'' density
wave correlations are a generic feature in lightly doped strongly correlated
antiferromagnetic materials, which may not necessarily be superconducting
(as evidenced in both Copper-Oxides and non-superconducting Nickel-Oxides
systems).

Moreover, there is preliminary evidence for stripes throughout the glassy
state region in $La_{2-x}Sr_{x}CuO_{4}$\cite{Yamada2000-LSCO-underdoped},
which extends into part of the superconducting phase. Thus, our analysis of
stripe disorder is relevant for understanding the general phase diagram of
High-Tc materials. (Though the role that stripes play in the particular
phenomenon of superconductivity is still controversial).

In all experimental systems, where the full charge \& spin stripes structure
was observed, it was found that both spin and charge order are glassy \cite%
{Tranquada99-(LaNd)SrCuO-glass,Hammel2000-LaEuSrCuO-glass,Hammel99-NiO-glass}%
. An intrinsic finite stripes correlation length signature of disorder is
consistently observed\cite{Tranquada99-(LaNd)SrCuO-glass,Cheong2000-NiO.133}%
. In this paper, based on our analysis of general types of disorder in
stripes, we are able to determine unambiguously the nature of the glassy
state in these systems. Our analysis is rather simple, based on elementary
general principles, and independent of model details and parameters.

The canonical two dimensional stripes structure comprise of hole rich lines
(charge stripes) which form anti-phase domain walls between
antiferromagnetic (AFM) correlated spin regions (see fig.2a). As a result,
the spin periodicity $a_{s}$ is twice the charge periodicity $a_{c}$. The
charge and spin correlations (in the periodic direction) are hence
inexorably connected. It is thus quite surprising that, experimentally, the
correlation lengths of charge and spin differ substantially; e.g., in
Copper-Oxides $(LaNd)_{2-x}Sr_{x}CuO_{4}$ (with $x>0.1$) the ratio of the
stripes spin correlation length $\xi _{s}$ to the charge correlation length $%
\xi _{c}$ is $\xi _{s}/\xi _{c}\approx 4/1$ \cite%
{Tranquada99-(LaNd)SrCuO-glass}.

We highlight a distinction between two ways by which disorder potential can
lead to finite correlation length. Conventional models attribute glassiness
to the existence of topological defects. In contrast, Giamarchi \& LeDoussal
recently argued for the existence of a thermodynamically distinct zero
temperature glassy state - labeled ''Bragg Glass'' - due purely to elastic
deformation\cite{BraggGlass}. In two dimensions it is reduced to a ''quasi
Bragg glass'' state, were disorder by elastic deformations result with
correlation length $\xi $ smaller than typical distance $R_{D}$ between
unpaired defects; $\xi \ll R_{D}$. Since the typical distinctions between
the two glassy states are not easily probed by experiments, the Bragg glass
state has been a subject of continuing controversy.

Kivelson\&Emery speculated that a stripes Bragg glass state\ may occur as
part of a general stripes phase diagram\cite{StripesLiquidCrystals}. In
stripes, due to the unique feature of having two distinct and coupled
periodic order parameters (spin and charge), we argue that the ratio of spin
and charge correlation lengths, $\xi _{s}/\xi _{c}$, provide a novel sharp
criterion for determining the dominant form of disorder in the system. We
conclude that all current experiments on stripes are compatible only with
the conjecture that stripe correlations are disordered primarily by
non-topological elastic deformations (see fig.2c) \cite{Note2,Note4}.

The line of argument is the following: We find that disorder dominated by
topological defects lead only to the possibility of having the charge
stripes correlation length larger than the spin stripes correlation length $%
\xi _{s}$, i.e., $\xi _{s}/\xi _{c}<1$. Consequently, we establish a robust
qualitative distinction between topological defects leading always to $\xi
_{s}/\xi _{c}<1$, while predominant elastic deformations leading always to $%
\xi _{s}/\xi _{c}>1$ (and {\em expectantly} $\xi _{s}/\xi _{c}=4$). On this
basis, we argue that the observation of $\xi _{s}/\xi _{c}\approx 4$ in $%
(LaNd)_{2-x}Sr_{x}CuO_{4}$ unequivocally indicates a quasi Bragg glass type
state. It implies that topological defects are much less relevant than
commonly assumed for glassy properties of Copper-Oxide systems\cite%
{Note2,Note4}. Expected spectral properties are discussed.

The charge stripes are modeled as classical elastic charge density wave
structures. Similarly, the AFM spin regions are discussed as a classical
two-dimensional Heisenberg model. Disorder is then incorporated as acting
directly on these classical objects. The observed correlation length should
not be confused with the size of the stripes domains, where stripes
orientation in distinct domains differ by 90$^{\circ }$. All of our analysis
concerns the disorder within a single stripes domain (i.e., all stripes have
the same general orientation).

\section{Disorder by topological defects}

The topology of the stripes charge lines is similar to that of smectic
liquid crystals \cite{StripesLiquidCrystals}. Since in this paper we are not
interested in significant changes of the stripes orientation, we ignore
disclination defects and focus on dislocations (fig.1). Though each
topological defect in the charge lines order induces also a topological
defect in the spin order, the respective disordering effects of the defects
are different.

The charge lines are antiphase domain walls of the spin order, i.e., the
local AFM order parameter undergoes a local $\pi $ phase slip across each
charge stripe line. Therefore, if a round trip goes through an odd number of
charge lines then it is enclosing a $\pi $ vortex of the spin order. Such is
the case for going around a line ending or a line splitting dislocation
defects, which are highlighted by colored circles in fig.1 (The defects are
caricaturistically drawn with rather sharp line angles, but more realistic
softening of the curves will not soften our conclusions). One should not
confuse dislocations in the charge stripes {\em lines} with any kind of
dislocations in the 2D Heisenberg spin lattice {\em points} (i.e., the
underlying Cu-O plane). The magnetic $\pi $-vortex defects are associated
purely with rotations in spin space. The resulting spin texture can be
either an XY model vortex (if spins remain confined to the plane), or
otherwise Skyrmeons in an $O\left( 3\right) $ model.

A connection with the more familiar spin glass model of
frustrated/non-frustrated plaquettes\cite{Plackets} in a 2D Heisenberg model
can be made by mapping onto a square lattice where, every bond which is cut
by a charge stripe is a ferromagnetic bond and otherwise an AFM bond.
Frustrated/non-frustrated plaquettes are those with an odd/even number of
AFM bonds respectively.

To recapitulate, every topological defect in the charge smectic order is
simultaneously associated with a topological defect in the spin order.
Therefore, it seems at first that disorder by topological defects would
inevitably lead to having the same charge and spin correlation lengths
(i.e., $\xi _{s}/\xi _{c}\approx 1$).

In fig.1, we show a possible arrangement of two dislocation defects (the
full 2D plane can be understood as composed of similar arrangements at a
given density of defects). The associated $\pi $-vortex spin defects are
known to limit the spin correlation length in the plane, yet the effect on
charge correlations is less trivial. While any dislocation defect in a
smectic topology destroys the correlation {\em along} the lines ($x$-axis in
fig.1), the effect on correlations {\em perpendicular} to the lines ($z$%
-axis in fig.1) is more subtle. The experimentally measured stripes
correlations which are discussed throughout this paper are the ones
perpendicular to the stripe lines.

To be precise, imagine an x-ray scattering experiment performed on the
''window'' of charge stripes configuration depicted in fig.1. If the beam's
spot size is narrow (e.g., like the dashed yellow line) then obviously the
observed peak width will not differ from that of perfectly ordered stripes
(within the size of the depicted stripes figure). If the beam's spot sized
is widened to size of the window of fig.1 then there will be additional
second harmonic satellite peaks, but the width of all peaks will still
remain that of the perfectly ordered stripes system (with an added weak
background due to the relatively small deformation regions).

In conclusion, due to the different character of the charge and spin order
parameters, the distance between topological defects limits the spin
correlation length but the charge stripes correlation length (in the
modulation direction) is less affected. Therefore, predominant topological
defects disorder always lead to having $\xi _{s}/\xi _{c}<1$.

\begin{figure}
\begin{center}
\leavevmode\epsfxsize=2in 
\epsfbox{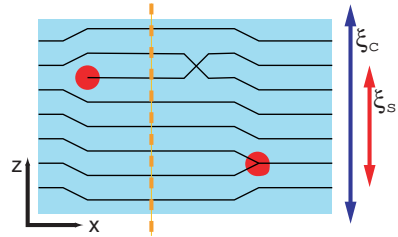} 
\end{center}
\caption{Possible configuration of two topological defects}
\label{fig-2}
\end{figure}

\section{Disorder by elastic deformations}

In the previous section we established the effect of topological defects. We
now examine the extreme opposite case, where topological defects are
excluded and only elastic deformations dominate the stripe's disorder
(fig.2c). i.e., a perfect Bragg glass stripes state. We show that, unlike
the case of disorder by topological defects, a Bragg glass stripe state {\em %
allows for} $\xi _{s}/\xi _{c}>1$, and moreover it is {\em expected} that $%
\xi _{s}/\xi _{c}=4$.

\subsection{Correlation length ratio $\protect\xi _{s}/\protect\xi _{c}$}

First, we elucidate how a finite correlation length arise due to static
elastic deformations. Elastic deformations of the stripes look like fig.2c.
For clarity, we elaborate an approximate simpler model of disorder (depicted
in fig.2b): Picture the charge stripes as an array of parallel rigid rods
(fig.2a) with period $a_{c}=u_{0}^{j+1}-u_{0}^{j}$ (in the absence of
disorder), and that the effect of external disorder potential is only to
locally perturb the stripes separation.

The disorder potential induces deviations $\Delta a_{c}^{j}$ in the local
separation between stripes; $u^{j+1}-u^{j}=a_{c}+\Delta a_{c}^{j}$, with
equal probability of local compression and extension, i.e., $\Delta
a_{c}^{j}=\pm \left| \Delta a_{c}\left( P\right) \right| $, and which may
have a distribution of variable size deviations $\left| \Delta a_{c}\left(
P\right) \right| $ with probability $P$. The typical local deformation may
be small, $\left| \Delta a_{c}\left( P\right) \right| \ll a_{c}$. Yet, when
statistically accumulated over a length of $N$ stripes, the average size of
fluctuation in the distance $u^{j+N}-u^{j}$ from its ordered state (where $%
u_{0}^{j+N}-u_{0}^{j}=Na_{c}$) grows typically like $\sqrt{N}\left| \Delta
a_{c}\right| $, where $\left| \Delta a_{c}\right| \equiv \sqrt{\left\langle
\left| \Delta a_{c}\left( P\right) \right| ^{2}\right\rangle }$.

The Bragg glass correlation length $\xi _{c}$ is the scale on which relative
displacements of the stripes from their ordered position becomes of order of
the bare stripes spacing $a_{c}$. In the ordered state (fig.2a), a Bragg
peak results from the constructive interference of reflections from each
stripe line (at the stripes wave-vector). Due to disorder, when the
accumulated deviation from the ordered state $\sum_{j=1}^{N_{a}}\Delta
a_{c}^{j}$ gets to be about half the bare periodicity (see fig.2b), the
constructive interference is lost which give rise to a finite peak width in
a scattering experiment by which the correlation length is defined. Thus, we
define 
\begin{equation}
\xi _{c}\equiv N_{a}a_{c}=\left( \frac{a_{c}}{2\left| \Delta a_{c}\right| }%
\right) ^{2}a_{c}  \label{EQ.c-length}
\end{equation}
where $N_{a}$ is given by the condition 
\begin{equation}
\frac{a_{c}}{2}\approx \left\langle \left| \sum_{j=1}^{N_{a}}\Delta
a_{c}^{j}\right| \right\rangle =\sqrt{N_{a}}\left| \Delta a_{c}\right|
\label{EQ.corr.condition-Na}
\end{equation}

In the relevant materials, the disorder potential (due mostly to the dopant $%
Sr$ charge impurities out of the stripe plane) couples directly only to the
charge stripes. What is the effect on the spin correlation length? The key
observation is that, in the stripe structure, the spin order deviations are
forced to follow the charge lines, since the charge lines are anti-phase
domain walls irrespective of their position (elsewhere\cite%
{Zachar2000-AntiphaseWall} we explicitly defend this statement). In other
words, the spin order is enslaved to the charge order and has no independent
intrinsic elastic stiffness (in real space motion, not in spin space
rotations). This is an essential outcome of the frustrated phase separation
view of stripes\cite{Note3}.

Therefore, for the resulting spin correlation length, it is only the spin
period $a_{s}$ that replace $a_{c}$ on the left side of (\ref%
{EQ.corr.condition-Na}) in the defining condition which relate the bare
period to the associated finite correlation length due to elastic
deformations, while the average disorder ''steps'' remain $\left| \Delta
a_{c}\right| $ per charge line also for the spin order. Thus, 
\begin{eqnarray}
\xi _{s} &=&\left( \frac{a_{s}}{2\left| \Delta a_{c}\right| }\right)
^{2}a_{c} \\
\xi _{s}/\xi _{c} &=&\left( \frac{a_{s}}{a_{c}}\right) ^{2}=4.
\end{eqnarray}%
Which is exactly the experimental observation\cite%
{Tranquada99-(LaNd)SrCuO-glass}, with no free adjustable parameters. The
above conclusion is insensitive to details of the proper elastic model of
stripes (and also insensitive to the characteristic deformation size $\left|
\Delta a_{c}\right| $). {\em It is the outcome of the generic stripes
structure where }$a_{s}/a_{c}=2${\em , and the generic origin of finite
correlation length due to random small deformations in periodic systems}.

\begin{figure}
\begin{center}
\leavevmode\epsfxsize=3in 
\epsfbox{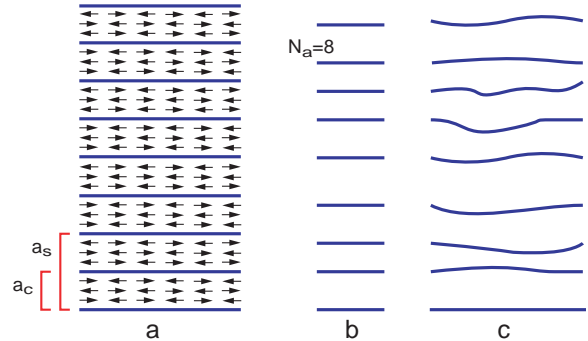} 
\end{center}
\caption{(a) Canonical stripes structure. 
(b) Model of elastic deformations of parallel stripes with disordered spacings.
(c) "Realistic" image of elastic stripes deformations with no topological defects.}
\label{fig-3}
\end{figure}

Once the correct framework for analyzing the glassy state is identified, we
are in position to derive various implications. For example, we estimate the
strength of disorder as characterized by the ratio $\left( \frac{va_{c}^{2}}{%
2\left| V_{q}\right| }\right) ^{-1}$, where $v$ is the elastic coefficient
associated with compressing or stretching the spacing of the stripes with
respect to their preferred spacing $a_{c}$, and $V_{q}$ is the disorder
potential strength.

Properly, the charge stripes should be described by some form of anisotropic
2D elastic model (and moreover coupled to an underlying lattice which breaks
the rotation symmetry in the plane). The full elastic model is very
complicated, but a qualitative estimate can be derived using the limit model
of stripes as a one dimensional CDW (as in fig.2b). Using the method of \cite%
{Disorder-FukuyamaLee78} for weak disorder, the ensuing correlation length $%
\xi _{c}$ is estimated, 
\begin{equation}
\frac{\xi _{c}}{a_{c}}=\left( \frac{va_{c}^{2}}{2\left| V_{q}\right| }%
\right) ^{2/3}.
\end{equation}%
For $\left( LaNd\right) _{7/8}Sr_{1/8}CuO_{4}$, where $a_{c}\approx 16$\AA\
and $\xi _{c}\approx 100$\AA , we find $\left( \frac{va_{c}^{2}}{2\left|
V_{q}\right| }\right) ^{-1}\approx 1/15$, which indicates that the
assumption of weak disorder is at least self-consistent.

\subsection{Charge {\em and} spin glassiness without topological defects?}

Since the disorder potential due to the dopant $Sr$ impurities couples only
to the charge stripes, it is not a-priori clear how a concomitant spin
disorder character can be obtained other than due to topological defects.

A glassy state is characterized by a wide distribution of activation energies%
\cite{Hammel2000-LaEuSrCuO-glass}. For the frequency dependent conductivity $%
\sigma \left( \omega \right) $ associated with various fluctuation modes of
the charge stripes density wave, we can use the calculations of \cite%
{Disorder-FukuyamaLee78}. $\sigma \left( \omega \right) $ has the general
form of a peak at energy $\omega _{0}\approx v/\xi _{c}$ and peak-width of
similar magnitude $\Delta \approx \omega _{0}$. $v/\xi _{c}$ can be
interpreted as the excitation of a fluctuation mode on the order of the
pinning localization length ''domain''. The peak shape is {\em not}
Gaussian, and has power-law tails\cite{Disorder-FukuyamaLee78}. In contrast
with the case of topological defects, there is no contribution to the glassy
spin characteristics from static non-topological deformations of the charge
stripes. Yet, spin dynamics is affected through its coupling to the charge
stripes {\em dynamics}. Though an explicit microscopic theory of spin-charge
coupling in stripes is still missing\cite{StripesLandau}, it is expected
that the distribution of charge fluctuation frequencies translates into a
distribution of local spin fluctuation energies (possibly with similar
characteristics). Thus, when topological defects are dilute, glassy spin
behavior may be dominated by the contribution of non-topological stripe
fluctuations.

\section{Conclusions}

To summarize, our main conclusion is the following: In stripes, disorder
purely by topological defects configurations lead to $\xi _{s}/\xi _{c}<1$.
Therefore, the seemingly minute detail of experimental finding that $\xi
_{s}/\xi _{c}\approx 4$ in $\left( LaNd\right) _{7/8}Sr_{1/8}CuO_{2}$ is
enough to exclude the possibility of disorder {\em dominated} by topological
defects. Moreover, we find that for disorder by pure elastic deformations it
is expected that $\xi _{s}/\xi _{c}=4$ quite insensitively to details of the
elastic stripe models. Thus, any observation of $1<\xi _{s}/\xi _{c}<4$
implies that there is a {\em relatively dilute concentration of topological
defects} (with typical distance $R_{D}$) where $\xi _{c}<R_{D}<4\xi _{c}$,
which cut the spin correlation length $\xi _{s}$ to be below its expected
value in the absence of defects.

Hence we proclaim that $\left( LaNd\right) _{7/8}Sr_{1/8}CuO_{2}$ at low
temperatures is a quasi Bragg glass type disordered system, in the sense
that it is disordered mainly by elastic deformations resulting in stripes
correlation length $\xi _{c}$ {\em much smaller} than the typical distance $%
R_{D}$ between unpaired topological defects\cite{Note2,Note4}, i.e., $4\xi
_{c}<R_{D}$. Thus, any theoretical approaches and interpretation of the
glassy properties of the known experimental stripe systems should not be
blindly done by fitting parameters to conventional theory based on
topological defects (dressed in stripes costume).

Our analysis can be applied, for example, to the effect of Zinc doping on
stripes. It is not clear whether the main effect is of local pinning of the
stripes or also as nucleation centers for stripe defects. We propose that an
examination of the ratio $\xi _{s}/\xi _{c}$ will give the answer.

Following our analysis, non-superconducting $\left( LaNd\right)
_{7/8}Sr_{1/8}CuO_{2}$ becomes the first doped Copper-Oxide material whose
electronic ground-state, including interactions and disorder, is delineated.
It is also one of the first strong evidence for realization of the
controversial quasi Bragg glass state\cite{BraggGlass}.

\begin{acknowledgement}
I thank Jose Torres, A. Nersesyan, J. Zaanen, and S. Kivelson for
discussions.
\end{acknowledgement}


\begin{references}
\bibitem{Tranquada99-(LaNd)SrCuO-glass} J.M. Tranquada, N. Ichikawa and S.
Uchida, {Phys. Rev. B} {\bf 59},14712 (1999).

\bibitem{Hammel2000-LaEuSrCuO-glass} N.J. Curro {\sl et al}., {Phys. Rev. Lett} 
{\bf 85},642 (2000).

cond-mat/9911268.

\bibitem{Hammel99-NiO-glass} Y. Yoshinari, P.C. Hammel, S-W. Cheong, Phys.
Rev. Lett. {\bf 82},3536 (1999).

\bibitem{Cheong2000-NiO.133} S.-H. Lee, S-W. Cheong, K. Yamada, C. Majkrzak,
cond-mat/0002348.

\bibitem{Yamada2000-LSCO-underdoped} M. Matsuda {\sl et al}.,
cond-mat/0003466.

\bibitem{FrustratedPhaseSeparation} V.~J.~Emery and S.~A.~Kivelson, Physica 
{\bf C 209}, 597 (1993); G. Seibold, C. Castellani, C. Di Castro, M. Grilli, 
{Phys. Rev. B} {\bf 58},13506 (1998).

\bibitem{BraggGlass} T. Giamarchi and P. Le Doussal, in ``{\sl Spin Glasses
and Random Fields}'', ed. A.P. Young, World Scientific (Singapore) 1998, p.
321.

\bibitem{Note2} Some topological defects surely do exist (as expected in two
dimensions \cite{BraggGlass}), since there exist distinct stripe domains
where the stripe orientation differs by 90$^{\circ }$.

\bibitem{Note4} The existence of paired topological defect ''dipoles'' with
size much smaller than $\xi _{c}$ is not excluded by our analysis, and is of
no consequence for our discussion.

\bibitem{StripesLiquidCrystals} S.A. Kivelson, E. Fradkin, and V.J. Emery, 
{\it Nature} {\bf 393},550 (1998); S.A. Kivelson and V.J. Emery,
cond-mat/9809082.

\bibitem{Plackets} J. Villain, J. Phys. C, {\bf 10}, 4793 (1977).

\bibitem{Note3} In the frustrated phase separation approach\cite%
{FrustratedPhaseSeparation}, and as demonstrated in general Landau theory of
stripes\cite{StripesLandau}, the periodic stripe structure is determined by
charge self-organization while spin order is only subsequently enslaved.
Such scenario is supported by the observation that charge order is always
established prior to spin order\cite{Cheong2000-NiO.133}.

\bibitem{StripesLandau} Spin-Charge coupling within a Landau theory was
considered in O.~Zachar, S.~A.~Kivelson, and V.~J.~Emery, {Phys. Rev. B} 
{\bf 57},1422 (1998).

\bibitem{Zachar2000-AntiphaseWall} Oron Zachar, cond-mat/0001217.

\bibitem{Disorder-FukuyamaLee78} H. Fukuyama and P.A. Lee, {Phys. Rev. B} 
{\bf 17}, 535 (1978).
\end{references}
\end{document}